\documentstyle[twoside,amssymb]{article}

\catcode`\@=11
\long\def\@makefntext#1{
\protect\noindent \hbox to 3.2pt {\hskip-.9pt  
$^{{\eightrm\@thefnmark}}$\hfil}#1\hfill}		

\def\thefootnote{\fnsymbol{footnote}}
\def\@makefnmark{\hbox to 0pt{$^{\@thefnmark}$\hss}}	
	
\def\ps@myheadings{\let\@mkboth\@gobbletwo
\def\@oddhead{\hbox{}
\rightmark\hfil\eightrm\thepage}   
\def\@oddfoot{}\def\@evenhead{\eightrm\thepage\hfil
\leftmark\hbox{}}\def\@evenfoot{}
\def\sectionmark##1{}\def\subsectionmark##1{}}

\oddsidemargin=\evensidemargin
\renewcommand{\thefootnote}{\fnsymbol{footnote}}

\newcounter{sectionc}\newcounter{subsectionc}\newcounter{subsubsectionc}
\renewcommand{\section}[1] {\vspace{12pt}\addtocounter{sectionc}{1} 
\setcounter{subsectionc}{0}\setcounter{subsubsectionc}{0}\noindent 
	{\tenbf\thesectionc. #1}\par\vspace{5pt}}
\renewcommand{\subsection}[1] {\vspace{12pt}\addtocounter{subsectionc}{1} 
	\setcounter{subsubsectionc}{0}\noindent 
	{\bf\thesectionc.\thesubsectionc. {\kern1pt \bfit #1}}\par\vspace{5pt}}
\renewcommand{\subsubsection}[1] {\vspace{12pt}\addtocounter{subsubsectionc}{1}
	\noindent{\tenrm\thesectionc.\thesubsectionc.\thesubsubsectionc.
	{\kern1pt \tenit #1}}\par\vspace{5pt}}

\newcounter{appendixc}
\newcounter{subappendixc}[appendixc]
\newcounter{subsubappendixc}[subappendixc]
\renewcommand{\thesubappendixc}{\Alph{appendixc}.\arabic{subappendixc}}
\renewcommand{\thesubsubappendixc}
	{\Alph{appendixc}.\arabic{subappendixc}.\arabic{subsubappendixc}}

\renewcommand{\appendix}[1] {\vspace{12pt}
        \refstepcounter{appendixc}
        \setcounter{figure}{0}
        \setcounter{table}{0}
        \setcounter{lemma}{0}
        \setcounter{theorem}{0}
        \setcounter{corollary}{0}
        \setcounter{definition}{0}
        \setcounter{equation}{0}
        \renewcommand{\thefigure}{\Alph{appendixc}.\arabic{figure}}
        \renewcommand{\thetable}{\Alph{appendixc}.\arabic{table}}
        \renewcommand{\theappendixc}{\Alph{appendixc}}
        \renewcommand{\thelemma}{\Alph{appendixc}.\arabic{lemma}}
        \renewcommand{\thetheorem}{\Alph{appendixc}.\arabic{theorem}}
        \renewcommand{\thedefinition}{\Alph{appendixc}.\arabic{definition}}
        \renewcommand{\thecorollary}{\Alph{appendixc}.\arabic{corollary}}
        \renewcommand{\theequation}{\Alph{appendixc}.\arabic{equation}}
        \noindent{\tenbf Appendix \theappendixc #1}\par\vspace{5pt}}
\newcommand{\subappendix}[1] {\vspace{12pt}
        \refstepcounter{subappendixc}
        \noindent{\bf Appendix \thesubappendixc. {\kern1pt \bfit #1}}
	\par\vspace{5pt}}
\newcommand{\subsubappendix}[1] {\vspace{12pt}
        \refstepcounter{subsubappendixc}
        \noindent{\rm Appendix \thesubsubappendixc. {\kern1pt \tenit #1}}
	\par\vspace{5pt}}

\newcommand{\textlineskip}{\baselineskip=13pt}
\newcommand{\smalllineskip}{\baselineskip=10pt}

\def\eightcirc{
\begin{picture}(0,0)
\put(4.4,1.8){\circle{6.5}}
\end{picture}}
\def\eightcopyright{\eightcirc\kern2.7pt\hbox{\eightrm c}} 

\newcommand{\copyrightheading}[1]
	{\vspace*{-2.5cm}\smalllineskip{\flushleft
	{\footnotesize Modern Physics Letters A, #1}\\
	{\footnotesize $\eightcopyright$\, World Scientific Publishing
	 Company}\\
	 }}

\newcommand{\publisher}[2]{{\begin{center}\footnotesize\smalllineskip 
	Received #1\\
	Revised #2
	\end{center}
	}}

\def\abstracts#1#2#3{{
	\centering{\begin{minipage}{4.5in}\footnotesize\baselineskip=10pt
	\parindent=0pt #1\par 
	\parindent=15pt #2\par
	\parindent=15pt #3
	\end{minipage}}\par}} 

\renewenvironment{thebibliography}[1]
	{\frenchspacing
	 \ninerm\baselineskip=11pt
	 \begin{list}{\arabic{enumi}.}
        {\usecounter{enumi}\setlength{\parsep}{0pt}     
	 \setlength{\leftmargin 12.7pt}{\rightmargin 0pt} 
         \setlength{\itemsep}{0pt} \settowidth
	{\labelwidth}{#1.}\sloppy}}{\end{list}}

\newcounter{itemlistc}
\newcounter{romanlistc}
\newcounter{alphlistc}
\newcounter{arabiclistc}

\newcommand{\fcaption}[1]{
        \refstepcounter{figure}
        \setbox\@tempboxa = \hbox{\footnotesize Fig.~\thefigure. #1}
        \ifdim \wd\@tempboxa > 5in
           {\begin{center}
        \parbox{5in}{\footnotesize\smalllineskip Fig.~\thefigure. #1}
            \end{center}}
        \else
             {\begin{center}
             {\footnotesize Fig.~\thefigure. #1}
              \end{center}}
        \fi}

\newcommand{\tcaption}[1]{
        \refstepcounter{table}
        \setbox\@tempboxa = \hbox{\footnotesize Table~\thetable. #1}
        \ifdim \wd\@tempboxa > 5in
           {\begin{center}
        \parbox{5in}{\footnotesize\smalllineskip Table~\thetable. #1}
            \end{center}}
        \else
             {\begin{center}
             {\footnotesize Table~\thetable. #1}
              \end{center}}
        \fi}

\def\@citex[#1]#2{\if@filesw\immediate\write\@auxout
	{\string\citation{#2}}\fi
\def\@citea{}\@cite{\@for\@citeb:=#2\do
	{\@citea\def\@citea{,}\@ifundefined
	{b@\@citeb}{{\bf ?}\@warning
	{Citation `\@citeb' on page \thepage \space undefined}}
	{\csname b@\@citeb\endcsname}}}{#1}}

\newif\if@cghi
\def\cite{\@cghitrue\@ifnextchar [{\@tempswatrue
	\@citex}{\@tempswafalse\@citex[]}}
\def\citelow{\@cghifalse\@ifnextchar [{\@tempswatrue
	\@citex}{\@tempswafalse\@citex[]}}
\def\@cite#1#2{{$\null^{#1}$\if@tempswa\typeout
	{IJCGA warning: optional citation argument 
	ignored: `#2'} \fi}}

\def\pmb#1{\setbox0=\hbox{#1}
	\kern-.025em\copy0\kern-\wd0
	\kern.05em\copy0\kern-\wd0
	\kern-.025em\raise.0433em\box0}

\def\fnt#1#2{\footnotetext{\kern-.3em
	{$^{\mbox{\scriptsize #1}}$}{#2}}}

\def\fpage#1{\begingroup
\voffset=.3in
\thispagestyle{empty}\begin{table}[b]\centerline{\footnotesize #1}
	\end{table}\endgroup}

\def\runninghead#1#2{\pagestyle{myheadings}
\markboth{{\protect\footnotesize\it{\quad #1}}\hfill}
{\hfill{\protect\footnotesize\it{#2\quad}}}}
\font\tenrm=cmr10
\font\tenit=cmti10 
\font\tenbf=cmbx10
\font\bfit=cmbxti10 at 10pt
\font\ninerm=cmr9

\font\eightrm=cmr8

\def\qed{\hbox{${\vcenter{\vbox{			
   \hrule height 0.4pt\hbox{\vrule width 0.4pt height 6pt
   \kern5pt\vrule width 0.4pt}\hrule height 0.4pt}}}$}}

\renewcommand{\thefootnote}{\fnsymbol{footnote}}	

\begin{document}
\setlength{\textheight}{7.7truein}  

\runninghead{Creation of Dirac particles in the presence of
$\ldots$}{Creation of Dirac particles in the presence of
$\ldots$}

\normalsize\textlineskip
\thispagestyle{empty}
\setcounter{page}{1}

\copyrightheading{}			

\vspace*{0.88truein}

\fpage{1}
\centerline{\bf CREATION OF DIRAC PARTICLES IN THE PRESENCE OF A}
\baselineskip=13pt
\centerline{\bf CONSTANT ELECTRIC FIELD IN AN ANISOTROPIC} 
\baselineskip=13pt
\centerline{\bf BIANCHI I UNIVERSE}
\vspace*{0.37truein}
\centerline{\footnotesize V\'ICTOR M. VILLALBA\footnote{On leave of
absence from Centro de F\'{\i}sica, Instituto Venezolano de Investigaciones
Cient\'{\i}ficas, IVIC, Apdo 21827, Caracas 1020-A, Venezuela.}
\footnote{e-mail: villalba@th.physik.uni-frankfurt.de}
\footnote{Alexander von Humboldt Fellow} \  and WALTER GREINER}

\baselineskip=12pt
\centerline{\footnotesize\it Institut f\"ur Theoretische Physik.
Universit\"at Frankfurt }
\baselineskip=10pt
\centerline{\footnotesize\it D-60054 Frankfurt am Main, Germany}
\vspace*{0.225truein}

\publisher{(received date)}{(revised date)}

\vspace*{0.21truein}
\abstracts{In this article we compute the density of Dirac particles created by
a cosmological anisotropic Bianchi I
universe in the presence of a constant electric field. We show that the
particle distribution becomes thermal when one neglects the electric
interaction.}{}{}



\vspace*{1pt}\textlineskip	

\setcounter{footnote}{0}
\renewcommand{\thefootnote}{\alph{footnote}}


\section{Introduction}

During the last decades a great effort has been made in understanding
quantum processes in strong fields. Quantum field theory in the presence of
strong fields is in general a theory associated with unstable vacua. The
vacuum instability leads to many interesting features, among them particle
creation is perhaps the most interesting non-perturbative phenomenon.
After the publication of the articles by Parker\cite{Parker,Parker2} and 
Zeldovich\cite{Zeldovich}, the study of quantum effects in cosmology became
a very active research field. In the last thirty years a large body of papers
has been published on the problem of vacuum effects in isotropic and homogeneous 
gravitational
backgrounds, mainly in deSitter and Robertson Walker models,  and only a few
try to discuss quantum processes in anisotropic Universes. 

 In order to analyze the mechanism of particle creation in cosmological
backgrounds we have at our disposal different techniques such as the
adiabatic approach\cite{Parker,Fulling}, the Feynman path integral method 
\cite{Chitre}, the Hamiltonian diagonalization technique\cite
{Grib,Bukhbinder}, as well as the Green function approach\cite{Gitman}

The study of quantum effects in gravitational backgrounds with initial
singularities presents an additional difficulty. The techniques commonly
applied in order to define particle states are based on the existence of a
timelike Killing vector or an asymptotically static metric\cite{Birrel}. A
different approach is needed to circumvent the problem related to the
initial singularity. In this direction, the Feynman path-integral method has
been applied to the quantization of a scalar field moving in the
Chitre-Hartle Universe\cite{Chitre,Fischetti}. This model has a curvature
singularity at $t=0$, and it is perhaps the best known example where a time
singularity appears and consequently any adiabatic prescription in order to
define particle states fails. A spin $1/2$ extension has been considered by
Sahni\cite{Sahni}.

A different approach to the problem of classifying single particle states on
curved spaces, is based on the idea of diagonalizing the Hamiltonian. This
technique permits one to compute the mean number of particles produced by a
singular cosmological model, and in particular by the Chitre-Hartle
Universe\cite{Chitre}.

An interesting scenario for discussing particle creation processes is the
anisotropic universe associated with the metric
\begin{equation}
ds^{2}=-dt^{2}+t^{2}(dx^{2}+dy^{2})+dz^{2}  \label{1}
\end{equation}
The line element (\ref{1}) presents a space-like singularity at $t=0.$ The
scalar curvature is $R=2/t^{2}$, and consequently, the adiabatic
approach\cite{Birrel} cannot be applied in order to define particle states.
Creation
of scalar particles in the anisotropic universe (\ref{1}) was originally
computed by Nariai with the help of Feynman propagators, obtaining as a
result that the creation process occurs in accordance with the black-body
thermal law in a 2-dimensional hypersurface related to the anisotropic cosmic
expansion.  With the
help of the Hamiltonian diagonalization method\cite{Grib,bukh2,bukh3},
Bukhbinder\cite{Bukhbinder} could compute the rate of scalar
particles produced in the space with the metric (\ref{1}), obtaining as a
result, a Bose-Einstein distribution. In Ref.\cite{Odintsov1} this result 
has been extended including a time dependent electric field  
More recently\cite{villalba,villalba2}, a
quasiclassical approach has been applied to compute the rate of scalar as
well as Dirac particles in the cosmological universe associated with the 
metric (\ref{1}).

The introduction of an external electric field permits one to consider an
additional source of quantum processes. The density of particles created by
an intense electric field was first calculated by Schwinger\cite{schwinger},
different authors\cite{Grib,Nikishov} have discussed this problem. Pair
creation of scalar particles by a constant electric field in a 2+1 de Sitter
cosmological universe has been analyzed by Garriga\cite{Garriga}. Quantum
effects associated with scalar and spinor particles in a quasi-Euclidean
cosmological model with a constant electric field are discussed by
Bukhbinder and Odintsov\cite{Odintsov}. 
It is worth mentioning
that the presence of primordial electric fields could enhance the particle
creation mechanism and also produces deviations from the thermal spectrum.

The purpose of the present article is to discuss the production of
Dirac particles in the anisotropic cosmological background associated with
the line element (\ref{1}) in the presence of a constant
electric field. In order to compute the rate of particles created we apply a
quasiclassical approach that has been successfully applied in different 
scenarios\cite
{villalba2,Lotze1,Lotze2,Villalba1}. The idea behind the method is the
following: First, we solve the relativistic Hamilton-Jacobi equation and,
looking at its solutions, we identify positive and negative frequency modes.
Second, we solve the Dirac equation and, after comparing
with the results obtained for the quasiclassical limit, we identify the
positive and negative frequency states.

The paper is structured as follows.
In Sec. 2 we solve the relativistic Hamilton-Jacobi equation and compute
the quasiclassical energy modes. In Sec. 3, after separating variables, 
we solve the Dirac equation and obtain the density of Dirac 
particles created. The discussion of the results and final remarks are presented
in Sec. 4.

\section{Hamilton-Jacobi equation}

The relativistic Hamilton-Jacobi equation can be written as 
\begin{equation}
g^{\alpha \beta }(\frac{\partial S}{\partial x^{\alpha }}-eA_{\alpha })(%
\frac{\partial S}{\partial x^{\beta }}-eA_{\beta })+m^{2}=0,  \label{2}
\end{equation}
where here and elsewhere we adopt the convention $c=1$ and $\hbar =1.$

The vector potential $A_{\alpha }$ 
\begin{equation}
A_{\alpha }=(0,0,0,-Et),
\end{equation}
corresponds to a constant electric field $E\hat{k}$. The corresponding
invariants $F^{\mu \upsilon }F_{\mu \nu }=-2E^{2}$ and $F^{\mu \upsilon \ast
}F_{\mu \nu }=0$ indicate that there is no magnetic field. Since the metric $%
g_{\alpha \beta }$ associated with the line element (\ref{1}) only depends
on $t,$ the function $S$ can be separated as 
\begin{equation}
S=F(t)+k_{x}x+k_{y}y+k_{z}z.  \label{3}
\end{equation}
Substituting (\ref{3}) into (\ref{2}) we obtain 
\begin{equation}
\dot{F}^{2}=\frac{k_{x}^{2}+k_{y}^{2}}{t^{2}}+(k_{z}+eEt)^{2}+m^{2}.
\label{4}
\end{equation}
The solution of Eq. (\ref{4}) presents the following asymptotic behavior: 
\begin{equation}
\lim_{t\rightarrow \infty }F=\pm \frac{1}{2}t\sqrt{e^{2}E^{2}t^{2}-m^{2}}\mp 
\frac{m^{2}}{2eE}\log (eEt+\sqrt{e^{2}E^{2}t^{2}-m^{2}}),  \label{5}
\end{equation}
\begin{equation}
\Phi =e^{iS}\rightarrow Ce^{\pm \frac{i}{2}eEt^{2}}(eEt)^{\mp \frac{im^{2}}{%
2eE}}  \label{fi}
\end{equation}
as $t\rightarrow \infty ,$ and 
\begin{equation}
\lim_{t\rightarrow 0}F=\pm \sqrt{(k_{x}^{2}+k_{y}^{2})\log t,}\ \Phi
=e^{iS}\rightarrow Ct^{\pm i\sqrt{k_{x}^{2}+k_{y}^{2}}},\   \label{6}
\end{equation}
as $t\rightarrow 0,$ that is, in the initial singularity. Notice that the
time dependence of the relativistic wave function is obtained via the
exponential operation $\Phi \rightarrow \exp (iS).$ Here the behavior of 
positive and negative frequency states is
selected depending on the sign of the operator $i\partial _{t}.$ Positive
frequency modes will have positive eigenvalues and for negative frequency
states we will have negative eigenvalues. Then in Eqs. (\ref{5}), (\ref{fi}%
) and (\ref{6}), upper signs are associated with negative frequency values and the lower
signs correspond to positive frequency states. 
After making this identification we proceed to
analyze the solutions of the Dirac equation in the cosmological
background (\ref{1}).

\section{Solution of the Dirac Equation}

The covariant Dirac equation in curved space in the presence of
electromagnetic fields can be written as follows 
\begin{equation}
\{\gamma ^{\mu }(\partial _{\mu }-\Gamma _{\mu }-ieA_{\mu })+m\}\Psi =0,
\label{Dir}
\end{equation}
where the curved gamma $\gamma ^{\mu }$ matrices satisfy the anticommutation
relation $\left\{ \gamma ^{\mu },\gamma ^{\nu }\right\}=2g^{\mu \nu }$
and the spinor connections $\Gamma _{\mu }$ are 
\begin{equation}
\Gamma _{\mu }=\frac{1}{4}g_{\lambda \alpha }\left[ \left( \frac{\partial
b_{\nu }^{\beta }}{\partial x^{\mu }}\right) a_{\beta }^{\alpha }-\Gamma
_{\nu \mu }^{\alpha }\right] s^{\lambda \nu },  \label{spinc}
\end{equation}
where 
\begin{equation}
s^{\lambda \nu }=\frac{1}{2}(\gamma ^{\lambda }\gamma ^{\nu }-\gamma ^{\nu
}\gamma ^{\lambda }).
\end{equation}
The matrices $b_{\alpha }^{\beta }$, $a_{\beta }^{\alpha }$ establish the
connection between the curved $\gamma ^{\mu }$ and Minkowski $\tilde{\gamma}%
^{\mu }$ Dirac matrices as follows 
\begin{equation}
\gamma _{\mu }=b_{\mu }^{\alpha }\tilde{\gamma}_{\alpha },\mbox{ }\gamma
^{\mu }=a_{\beta }^{\mu }\tilde{\gamma}^{\beta }
\end{equation}
and 
\begin{equation}
\tilde{\gamma}^{\lambda }\tilde{\gamma}^{\nu }+\tilde{\gamma}^{\nu }\tilde{%
\gamma}^{\lambda }=2\eta ^{\lambda \nu }
\end{equation}
Since the line element (\ref{1}) is diagonal, we choose to work in the
diagonal tetrad 
\begin{equation}
\gamma ^{\mu }=\sqrt{\left| g^{\mu \mu }\right| }\tilde{\gamma}^{\mu },\ \ %
\mbox{no sum}.  \label{conec}
\end{equation}
Substituting (\ref{conec}) into (\ref{spinc}), we obtain 
\begin{equation}
\Gamma _{1}=\frac{1}{2}\tilde{\gamma}^{0}\tilde{\gamma}^{1},\ \Gamma _{2}=%
\frac{1}{2}\tilde{\gamma}^{0}\tilde{\gamma}^{2},\ \Gamma _{3}=0,\ \Gamma
_{0}=0  \label{con}
\end{equation}
and the Dirac equation (\ref{Dir}) takes the form 
\begin{equation}
\left\{ \tilde{\gamma}^{0}\frac{\partial }{\partial t}+\frac{1}{t}(\tilde{%
\gamma}^{1}\frac{\partial }{\partial x}+\tilde{\gamma}^{2}\frac{\partial }{%
\partial y})+\tilde{\gamma}^{3}(\frac{\partial }{\partial z}+ieEt)+m\right\}
\Psi _{0}=0,  \label{Dira}
\end{equation}
where $\Psi _{0}=t\Psi .$ The factor $t$ was introduced in order to cancel
the contribution due to the spinor connections \ (\ref{con}). The equation (%
\ref{Dira}) can be written as a sum of two first order commuting
differential operators as follows\cite{Villalba3,Villalba4} 
\begin{equation}
(\hat{K}_{1}+\hat{K}_{2})\Phi =0
\end{equation}
\begin{equation}
\hat{K}_{2}\Phi ={\emph k}\Phi =-\hat{K}_{1}\Phi ,
\end{equation}
where the spinor $\Phi $ is related to $\Psi _{0}$ via the equation
\begin{equation}
\tilde{\gamma}^{3}\tilde{\gamma}^{0}\Psi _{0}=\Phi ,\ 
\end{equation}
and ${\emph k}$ is a separation constant. The operators $\hat{K}_{1}$ and $\ 
\hat{K}_{2}$ read 
\begin{equation}
\hat{K}_{1}=t\left[ \gamma ^{3}\frac{\partial }{\partial t}+\gamma ^{0}(%
\frac{\partial }{\partial z}-ieEt)+m\gamma ^{3}\gamma ^{0}\right] 
\label{k1}
\end{equation}
\begin{equation}
\hat{K}_{2}=\left( \tilde{\gamma}^{1}\frac{\partial }{\partial x}+\tilde{%
\gamma}^{2}\frac{\partial }{\partial y}\right) \tilde{\gamma}^{3}\tilde{%
\gamma}^{0}.  \label{K2}
\end{equation}
Since Eq. (\ref{Dira}) commutes with $-i\nabla $, the spinor $\Phi $ can be
written as $\Phi =\Phi _{0}\exp (i(k_{x}x+k_{y}y+k_{z}z))$. Choosing to
work in the following representation of the Dirac matrices 
\begin{equation}
\tilde{\gamma}^{0}=\left( 
\begin{array}{cc}
-i\sigma ^{1} & 0 \\ 
0 & i\sigma ^{1}
\end{array}
\right) ,\ \tilde{\gamma}^{1}=\left( 
\begin{array}{cc}
0 & i \\ 
-i & 0
\end{array}
\right) \ 
\tilde{\gamma}^{2}=\left(
\begin{array}{cc}
\sigma ^{2} & 0 \\ 
0 & -\sigma ^{2}
\end{array}
\right) \ \tilde{\gamma}^{3}=\left( 
\begin{array}{cc}
\sigma ^{3} & 0 \\ 
0 & -\sigma ^{3}
\end{array}
\right)   \label{rep}
\end{equation}
we obtain that $\hat{K}_{2}\Phi ={\emph k}\Phi $ reduces to an algebraic equation
that permits one determine the relation between the components of the bispinor 
$\Phi $%
\begin{equation}
\Phi _{0}=\left( 
\begin{array}{c}
\Phi _{1} \\ 
\Phi _{2}
\end{array}
\right) =\left( 
\begin{array}{c}
\Phi _{1} \\ 
\frac{k_{x}\sigma ^{2}}{ik_{y}-{\emph k}}\Phi _{1}
\end{array}
\right),   \label{bisp}
\end{equation}
where eigenvalue $k$ is 
\begin{equation}
{\emph k}=i\sqrt{k_{x}^{2}+k_{y}^{2}}.
\end{equation}
Using the representation (\ref{rep}) and taking into account the spinor
structure (\ref{bisp}) we obtain that, for $k_{z}=0$,  Eq. (\ref{k1}) reduces
to the system of equations
\begin{equation}
\label{e1}
\frac{d\varphi _{1}}{dt}+\frac{k}{t}\varphi _{1}+(m+ieEt)\varphi _{2}=0,
\end{equation}
\begin{equation}
\label{e2}
-\frac{d\varphi _{2}}{dt}+\frac{k}{t}\varphi _{2}+(m-ieEt)\varphi _{1}=0.
\end{equation}
In this way, we have reduced the problem of solving \ Eq. (\ref{Dira}) to
that of finding the solution of Eqs. (\ref{e1})-(\ref{e2}).
From  (\ref{e1}) and (\ref{e2}) we obtain the following
second order differential equation 
\begin{equation}
\frac{d^{2}\psi _{2}}{dt^{2}}+\left( -\frac{k^{2}-2k+3/4}{t^{2}}%
+m^{2}+e^{2}E^{2}t^{2}\right) \psi _{2}=0,  \label{segu}
\end{equation}
where we have introduced the  variable $\psi _{2}=t^{-1/2}\phi _{2}$ and
have neglected the mass  in the first-order variation of $\phi _{2}$. The
solution of (\ref{segu}) can be expressed in terms of Whittaker functions as 
\begin{equation}
\psi _{2}=C_{1}M_{\lambda ,\mu }(ieEt^{2})+C_{2}W_{\lambda
,\mu }(ieEt^{2}),
\end{equation}
where 
\begin{equation}
\lambda =-\frac{im^{2}}{4eE},\ \mu =\frac{k}{2}-\frac{1}{2},
\end{equation}
and $C_{1}$ and $C_{2}$ are arbitrary constants. Analogously, \ after
introducing the new variable  $\psi _{1}=t^{-1/2}\phi _{1}$, we have
that the equation for $\psi _{1}$ is   
\begin{equation}
\frac{d^{2}\psi _{1}}{dt^{2}}+\left( -\frac{k^{2}+2k+3/4}{t^{2}}%
+m^{2}+e^{2}E^{2}t^{2}\right) \psi _{1}=0
\end{equation}
with solutions 
\begin{equation}
\psi _{1}=C_{3}M_{\lambda ,\mu +1}(ieEt^{2})+C_{4}W_{\lambda ,\mu
+1}(ieEt^{2})
\end{equation}
where the Whittaker functions $W_{k,\mu}(z)$ and $M_{k,\mu}(z)$ can be expressed
in terms of confluent hypergeometric fuctions\cite{Abramowitz} as follows
\begin{equation}
M_{k,\mu }(z)=e^{-z/2}z^{1/2+\mu }M(\frac{1}{2}+\mu -k,1+2\mu ,z),
\label{Ku}
\end{equation}
\begin{equation}
W_{k,\mu }(z)=e^{-z/2}z^{1/2+\mu }W(\frac{1}{2}+\mu -k,1+2\mu ,z).
\label{Ku2}
\end{equation}
In order to construct the positive and negative frequency modes we use the asymptotic
behavior of the hypergeometric functions (\ref{Ku}) and (\ref{Ku2}) and compare the
solutions of Eq. (\ref{segu}) with (\ref{5}) and (\ref{6}), obtained after solving the
Hamilton-Jacobi equation. Using this procedure and looking at the asymptotic behavior
of $M_{k,\mu }(z)$ as $z\rightarrow 0$   
\begin{equation}
M_{k,\mu }(z)\backsim e^{-z/2}z^{1/2+\mu }  \label{M}
\end{equation}
and using the fact that all the coefficients in Eq. (\ref{segu}) are real, we obtain
that the positive and negative frequency solutions as $t\rightarrow 0$ are 

\begin{equation}
\label{delta0}
 \psi _{0}^{+}={\mathfrak {C}_{0}}^{+}M_{\lambda,\mu}(ieEt^2),\ \psi _{0}^{-}=(%
{\mathfrak {C}_{0}}^{+}M_{\lambda,\mu}(ieEt^2))^{\ast }={\mathfrak {C}_{0}}^{+}(-1)^{-%
\bar{\mu}+1/2}M_{\lambda,-\mu}(ieEt^2),
\end{equation}
where ${\mathfrak {C}_{0}^{+}}$ is a normalization constant. Analogously, \
looking at the behavior of $W_{k,\mu }(z)$ as $\left| z\right| \rightarrow
\infty $%
\begin{equation}
W_{k,\mu }(z)\backsim e^{-z/2}z^{k},  \label{W}
\end{equation}
we have that the corresponding positive and negative frequency modes as $%
 t \rightarrow +\infty $ are 
\begin{equation}
\label{infi}
\psi _{\infty }^{+}={\mathfrak {C}}_{\infty }^{+}W_{k,\mu}(ieEt^2 ),\
\psi _{\infty }^{-}={\mathfrak {C}}_{\infty }^{-}W_{-k,\mu}(-ieEt^2 ),
\end{equation}
where ${\mathfrak {C}_{\infty }^{+}}$ and ${\mathfrak {C}_{\infty }^{-}}$ are
normalization constants. 

Using the asymptotic expressions 

\begin{equation}
\lim_{a\rightarrow \infty }M(a,b,-z/a)/\Gamma (b)=z^{\frac{1}{2}-\frac{b}{2}%
}J_{b-1}(2\sqrt{z})
\end{equation}
\begin{equation}
\lim_{a\rightarrow \infty }U(a,b,-z/a)\Gamma (1+a-b)=\mp i\pi e^{\pi ib}z^{%
\frac{1}{2}-\frac{b}{2}}H_{b-1}^{(1,2)}(2\sqrt{z})
\end{equation}
we obtain that, as $E \rightarrow 0$, the positive and negative frequency modes 
(\ref{delta0}) and (\ref{infi}) reduce to those obtained through using the diagonalization method
\cite{Bukhbinder} as well as with the help of the semiclassical approach\cite{villalba}

The positive frequency mode as $t \rightarrow 0$. 
can be expressed in terms of the positive \ $\psi
_{\infty}^{+}$ and negative $\psi _{\infty}^{-}$ frequency modes via the Bogoliubov
transformation 
\begin{equation}
\psi _{0 }^{+}=\alpha \psi _{\infty}^{+}+\beta \psi _{\infty}^{-}
\end{equation}
The Whittaker function $M_{k,\mu }(z)$ can be expressed in terms of $%
W_{k,\mu }(z)$ as\cite{Abramowitz} 
\begin{equation}
 M_{k,\mu }(z)=\frac{\Gamma (2\mu +1)}{\Gamma (\mu -k+\frac{1}{2})}e^{-i\pi
k}W_{-k,\mu }(-z)+\frac{\Gamma (2\mu +1)}{\Gamma (\mu +k+\frac{1}{2})}%
e^{-i\pi (k-\mu -\frac{1}{2})}W_{k,\mu }(z).  \label{recu}
\end{equation}
Using the expression (\ref{recu}), we have that the negative frequency
solution $\psi _{0}^{-}$ can be written in terms of $\psi _{\infty }^{+}$
and $\psi _{\infty }^{-}$ as follows 
\begin{equation}
 \psi _{0}^{-}=\frac{\Gamma (2\mu +1)}{\Gamma (\mu -\lambda+\frac{1}{2})}e^{-i\pi
k}\psi _{\infty }^{-}+\frac{\Gamma (2\mu +1)}{\Gamma (\frac{1}{2}+\mu +\lambda)}%
(-1)^{-1/4}e^{-i\pi \mu (\lambda-\mu -\frac{1}{2})}(\psi _{\infty }^{-})^{\ast
},
\label{relation}
\end{equation}
where we have made use of the property $W_{-k,\mu }(-z)=(W_{k,\mu
}(z))^{\ast }$

Since we have been able to obtain single particle states for in the
vicinity of $t=0$ as well as in the asymptote $t\rightarrow \infty $, we can
compute the density of particles created by the gravitational field. With
the help of the Bogoliubov coefficients\cite{Grib,Birrel}. From (\ref
{relation}) and using the fact that $\psi_{0}^{-}=\alpha \psi _{\infty
}^{-}+\beta (\psi _{\infty }^{-})^{\ast }$, we obtain, 
\begin{equation}
\frac{\left| \beta \right| ^{2}}{\left| \alpha \right| ^{2}}=e^{2i\pi \mu }%
\frac{\left| \Gamma (\frac{1}{2}+\mu -\lambda)\right| ^{2}}{\left| \Gamma (\frac{1%
}{2}+\mu +\lambda)\right| ^{2}}.  \label{beta}
\label{rela}
\end{equation}
Substituting into (\ref{beta}) the values for $\mu $ and $k$ we obtain 
\begin{equation}
\label{div}
\frac{\left| \beta \right| ^{2}}{\left| \alpha \right| ^{2}}=e^{-\pi
k_{\perp }}\frac{\left( \frac{k_{\perp }}{2}-\frac{m^{2}}{4eE}\right) \sinh (%
\frac{\pi k_{\perp }}{2}-\frac{\pi m^{2}}{4eE})}{\left( \frac{k_{\perp }}{2}+%
\frac{m^{2}}{4eE}\right) \sinh (\frac{\pi k_{\perp }}{2}+\frac{\pi m^{2}}{4eE%
})},
\end{equation}
where we have used the relation\cite{Lebedev} 
\begin{equation}
\left| \Gamma (iy)\right| ^{2}=\frac{\pi }{y\sinh \pi y}
\end{equation}
The computation of the density of particles created is straightforward from (%
\ref{div}) and the normalization condition\cite{Mishima} of the wave
function 
\begin{equation}
\left| \alpha \right| ^{2}+\left| \beta \right| ^{2}=1,  \label{norma}
\end{equation}
then 
\begin{equation}
\label{ened}
n=\left| \beta \right| ^{2}=\left[ \left( \frac{\left| \beta \right| ^{2}}{%
\left| \alpha \right| ^{2}}\right) ^{-1}+1\right] ^{-1}.
\end{equation}

It is worth mentioning that, thanks to the relations (\ref{rela}) and (\ref
{norma}) we did not have to compute the normalization constants ${\mathfrak {C}%
_{\infty }^{+}}$, ${\mathfrak {C}_{\infty }^{-}}$ and ${\mathfrak {C}_{0}^{+}}$. \ \

\section{Results and Discussion} 
Let us analyze the
asymptotic behavior of \ (\ref{ened}) when the electric field vanishes.
Taking into account that $sinh(z)\sim \Theta(z)e^{\left| z\right| }/2$ as $%
z\rightarrow \infty $, we readily obtain
\begin{equation}
n\approx \frac{\left| \beta \right| ^{2}}{\left| \alpha \right| ^{2}}=\exp
(-2\pi k_{\perp}),  \label{espectro}
\end{equation}
which is the result obtained in\cite{villalba}. \ Expression (\ref{espectro}%
) \ corresponds to a two dimensional Fermi-Dirac thermal distribution.
In the case of strong electric fields the density number of scalar
particles created takes the form 
\begin{equation}
n\approx e^{-\pi k_{\perp } -\frac{\pi m^{2}}{2eE}},
\label{electrico}
\end{equation}
showing \ that the density of particles created by the cosmological
background and the electric field \ (\ref{electrico}) is\ a Fermi
distribution with \ a chemical potential proportional to $\frac{m^{2}}{eE}.$ 
 Integrating the particle density $n$ \ (\ref{electrico}) on momentum we
obtain the total number of particles created per unit volume. 
\begin{equation}
N=\frac{1}{V}\int ndk_{x}dk_{y}dk_{z}=\frac{1}{t^{2}(2\pi )^{2}}\int
nk_{\perp }dk_{\perp }dk_{z}.  \label{ene}
\end{equation}
In order to carry out the integration we have to notice that $n$ does not
depend on $k_{z}$ and consequently integration over $k_{z}$ is equivalent to
the substitution\cite{Grib,Nikishov} \ $\int dk_{z}\rightarrow eET$, where $T$
is the time of interaction of the external field. 
Substituting (\ref{electrico}) \ into (\ref{ene}) , we obtain that the
total number $N$ of particles per unit volume takes the form 
\begin{equation}
\label{numero}
N\approx \frac{eE}{4\pi ^{4}T}e^{-\frac{\pi m^{2}}{2eE}}
\end{equation}

Result (\ref{numero}) resembles the number of
particles created by a constant electric field in a Minkowski space\cite
{Grib,Nikishov}. It is worth mentioning that the number $N$ of particles per
unit volume is inversely proportional to $T^{-1}$ and vanishes as $%
T\rightarrow \infty $. The \ volume expansion of the anisotropic universe (%
\ref{1}) is faster than the particle creation process, therefore $N$
becomes negligible for large values of $T.$  Since the gravitational density
$\rho=1/(8\pi t^2)$, decreases faster than (\ref{numero}), the particle
creation mechanism effectively isotropizes in the anisotropic universe (\ref{1})
in the presence of strong electric fields. 

The results (\ref{electrico}), and (\ref{numero}) show that the
anisotropic cosmological background (\ref{1}), as well the constant
electric field, contribute to the creation of scalar particles. The
quasiclassical method gives a recipe for obtaining the positive and negative
frequency modes even when spacetime is not static and an external source is
present. The presence of the anisotropy with a constant electric field gives
rise to a particle distribution that is thermal only in the asymptotic
field regime. The method and results presented in this paper could be of
help to discuss quantum effects in more realistic anisotropic cosmological
scenarios.

\section{Acknowledments}{The authors wish to thank A. Mishra for valuable comments. 
One of the authors (V.M.V) acknowledges a fellowship from the
Alexander von Humboldt Stiftung}


\begin{thebibliography}{99}

\bibitem{Parker}L. Parker, {\textit Phys. Rev. Lett.} {\bf21},  562 (1968).
\bibitem{Parker2}L. Parker, {\it Asymptotic Structure of Space-Time} (Plenum, 1977)
p. 107.
\bibitem{Zeldovich}Ya. Zeldovich, and A. A. Starobinskii, {\it Sov. Phys.
JETP} {\bf34}, 1159 (1971).

\bibitem{Fulling}S. A. Fulling, {\textit Aspects of Quantum Field Theory in
Curved Space-Time} (Cambridge, 1991)

\bibitem{Chitre}D. M. Chitre and J. B. Hartle, {\it Phys. Rev D.} {\bf%
16,} 251 (1977).

\bibitem{Grib}A. A. Grib,  S. G. Mamaev V. M. Mostepanenko, {\it 
Quantum Vacuum effects in strong fields} (Energoatomizdat, 1988).


\bibitem{Bukhbinder}I. L. Bukhbinder, {\it Izv. Vuzov. Fizika} {\bf7,
} 3 (1980).

\bibitem{Gitman}S. Gavrilov, D. M. Gitman and S. D. Odintsov, {\it %
Int. J. Mod. Phys. A} {\bf12}, 4837 (1997).

\bibitem{Birrel}N. D. Birrel and P. C. W. Davies, {\it Quantum Fields
in Curved Space} (Cambridge, 1982).

\bibitem{Nariai}H. Nariai, {\it Prog. Theor. Phys.} \textbf{59}, 1532 (1978).

\bibitem{Fischetti}M. V. Fischetti, J. B. Hartle and B. L. Hu, {\it %
Phys. Rev. D} {\bf20}, 1757 (1979).

\bibitem{Sahni}V. Sahni, {\it Class. Quantum Grav.} {\bf1}, 579 (1984).

\bibitem{bukh2}I. L. Bukhbinder and D. M. Gitman, {\it Izv. Vuzov Fizika%
} {\bf3 }, 90 (1979).

\bibitem{bukh3}I. L. Bukhbinder, {\it Izv. Vuzov Fizika} {\bf 4}, 55 (1979)

\bibitem{Odintsov1} I. L. Bukhbinder and S. D. Odintsov, {\it Izv.
Vuzov. Fizika} {\bf5}, 93 (1983).

\bibitem{villalba} V. M. Villalba, {\it Int. J. Theor. Phys.} {\bf36},
1321 (1997).

\bibitem{villalba2} V. M. Villalba, {\it Phys. Rev D.} {\bf 52}, 3742 (1995).

\bibitem{schwinger}J. Schwinger, {\it Phys. Rev.} {\bf82} 664 (1951).

\bibitem{Nikishov}A. I. Nikishov, {\it Trudy FIAN No. 111} (Nauka,
1979).

\bibitem{Garriga}J. Garriga, {\it Phys. Rev. D}, 49, {\bf 6343} (1994).

\bibitem{Odintsov}I. L. Bukhbinder and S. D. Odintsov, {\it Izv. Vuzov.
Fizika} {\bf5}, 12 (1982).

\bibitem{Lotze1}K. H. Lotze, {\it Class. Quantum Grav.} {\bf2}, 351 (1985).

\bibitem{Lotze2}K. H. Lotze, {\it Class. Quantum Grav.} {\bf2}, 363 (1985).

\bibitem{Villalba1}V. M. Villalba {\it Phys. Rev D.} {\bf 60},
127501 (1999).

\bibitem{Villalba3}G. V. Shishkin and V. M. Villalba, {\it J. Math. Phys}
{\bf30}, 2132 (1989)

\bibitem{Villalba4} V. M. Villalba {\it Mod. Phys. Lett A.} {\bf8},
2351 (1993).

\bibitem{Abramowitz}M. Abramowitz and I. Stegun {\textit Handbook of
Mathematical Functions} (Dover, 1974).

\bibitem{Lebedev}N. N. Lebedev {\textit Special functions and their
applications} (Dover, 1972).

\bibitem{Mishima} T. Mishima and A. Nakayama A,  {\textit Phys. Rev D.} {\bf37%
} 348 (1988).


\end{thebibliography}
\end{document}